\documentclass{iaus}
\usepackage{graphicx}
\usepackage{amsmath}
\usepackage{natbib}
\usepackage{bm}
\graphicspath{{./fig/}{./png/}}

\title[The negative magnetic pressure effect]
{The negative magnetic pressure effect in stratified turbulence}

\author[K. Kemel et al.]
{K. Kemel$^{1,2}$, A. Brandenburg$^{1,2}$,
N. Kleeorin$^3$, and I. Rogachevskii$^3$}

\affiliation{$^1$NORDITA, AlbaNova University Center, Roslagstullsbacken 23, SE-10691 Stockholm, Sweden \\
$^2$Department of Astronomy, Stockholm University, SE--10691 Stockholm, Sweden\\
$^3$Department of Mechanical Engineering, Ben-Gurion University of the Negev,\\
POB 653, Beer-Sheva 84105, Israel}

\pubyear{2010}
\volume{273}
\pagerange{}
\jname{Physics of Sun and Star Spots}
\editors{}

 \def\simle{\mathrel{\hbox{\rlap{\hbox{\lower4pt\hbox{$\sim$}}}\hbox{$<$}}}}
 \def\simgr{\mathrel{\hbox{\rlap{\hbox{\lower4pt\hbox{$\sim$}}}\hbox{$>$}}}}

 \def\c2{^{12}{\rm C}}
 \def\c3{^{13}{\rm C}}
 \def\n14{^{14}{\rm N}}
 \def\c1213{^{12}{\rm C}/^{13}{\rm C}}
 \def\he3he4{^3{\rm He}/^4{\rm He}}

\newcommand{\SSSS}{\mbox{\boldmath ${\sf S}$} {}}
\newcommand{\EQ}{\begin{equation}}
\newcommand{\EE}{\end{equation}}
\newcommand{\EQA}{\begin{eqnarray}}
\newcommand{\EEA}{\end{eqnarray}}

\def\Pm{\mbox{\rm Pr}_M}

\begin{document}

\maketitle

\begin{abstract}

While the rising flux tube paradigm is an elegant theory, its basic
assumptions, thin flux tubes at the bottom of the convection zone
with field strengths two orders of magnitude above equipartition,
remain numerically unverified at best.
As such, in recent years the idea of a formation of sunspots near the top
of the convection zone has generated some interest.
The presence of turbulence can strongly enhance diffusive transport
mechanisms, leading to an effective transport coefficient formalism in
the mean-field formulation.
The question is what happens to these coefficients when the turbulence
becomes anisotropic due to a strong large-scale mean magnetic field.
It has been noted in the past that this anisotropy can also lead to highly non-diffusive behaviour.
In the present work we investigate the
formation of large-scale magnetic structures as a result of
a negative contribution of turbulence to the large-scale
effective magnetic pressure in the presence of stratification.
In direct numerical simulations of forced turbulence in a stratified box,
we verify the existence of this effect.
This phenomenon can cause formation of large-scale magnetic structures
even from initially uniform large-scale magnetic field.

\keywords{turbulence, MHD, sunspots}
\end{abstract}

\firstsection
\section{Introduction}

The standard explanation for the appearance of strong magnetic fields at the solar surface
involves the coherent rise of a tachocline-generated magnetic flux tube through
the solar convection zone.
Flux tube emergence simulations do give very promising results \citep{Remp09}, the paradigm looks elegant and is very textbook friendly,
but some of its assumptions are problematic: the integrity of flux tubes,
their rise, and even their very existence.

So far, numerical simulations have failed to produce the assumed thin
magnetic flux tubes in the tachocline \citep{Cat06,Par09}.
Magnetic buoyancy as the driving transport mechanism through the
convection zone can be dominated by downward pumping \citep{Nor92,Tob98}.
For the tubes to remain intact, strongly super-equipartition field
strengths and strong twists are required \citep{Fan01}.

Thus, it seems not unreasonable to explore alternative mechanisms of
formation of strong magnetic fields at the solar surface.
A number of positive arguments has been put forward \citep{B05} in favour
of formation of sunspots from local flux
concentrations near the surface.
In mean-field models, such magnetic instabilities have been
produced by introducing a magnetic dependence of the thermal eddy diffusivity
\citep{KM00} and the viscous stress tensor \citep{BKR10}.

Turbulence generally is associated with enhanced transport effects.
However, it also exhibits non-diffusive behaviour, generating
magnetic field on much larger scales than the driving scale.
An example is
turbulent dynamos that are able to produce large-scale
magnetic fields \citep{BS05}.
In a stratified layer, magnetic fields tend to become buoyantly unstable:
assuming a constant temperature across the magnetic flux tube,
pressure balance implies lower densities in regions of stronger magnetic field.
Now one can wonder what the role of turbulent pressure is in such an
equilibrium if the magnetic structures extend over several turbulent eddies.
The turbulent pressure associated with the convective fluid motions is
certainly not negligible and is strongly
affected by the background magnetic field.
The latter can be seen by evaluating the total turbulent dynamic pressure, here given for the isotropic case:
\begin{displaymath}
P_{\rm turb} = \textstyle{\frac{1}{3}}\overline{\rho u^2} + \textstyle{\frac{1}{6}}\overline{b^2}/\mu_0,
\end{displaymath}
here $\bm{u}$ and $\bm{b}$ are the velocity and magnetic fluctuations,
respectively, $\mu_0$ the vacuum permeability and $\rho$
the fluid density. Overbars indicate ensemble averaging.
As shown in direct numerical simulations \citep[][hereafter referred to as BKR]{BKR10},
in forced turbulence with imposed uniform large-scale magnetic field,
the total turbulent energy is approximately conserved in this parameter regime.
\begin{displaymath}
\textstyle{\frac{1}{2}}\overline{\rho u^2} + \textstyle{\frac{1}{2}}\overline{b^2}/\mu_0 \equiv E_{\rm tot}\approx\mbox{const}.
\end{displaymath}
As a result, one finds a reversed feedback from the magnetic fluctuations on the turbulent pressure:
\begin{displaymath}
P_{\rm turb} = -\textstyle{\frac{1}{6}}\overline{b^2}/\mu_0 +2E_{\rm tot} /3
\end{displaymath}
\citep[][hereafter referred to as RK07]{KRR90,RK07}.
One can see that the effective mean magnetic pressure force is reduced
and can, in a certain parameter
range, be reversed.
This effect would then counteract the aforementioned buoyancy instability
in the presence of turbulence.
RK07 have suggested that the reversed feedback instability
could lead to the formation of magnetic flux concentrations near the
solar surface.

Mean-field magnetohydromagnetic simulations by BKR confirmed the basic
pheno\-me\-non of magnetic flux concentration
by the effect of turbulence on the mean Lorentz force and for sufficient stratification, a linear instability was found.
Direct numerical simulations (DNS) by BKR confirmed the reversed feedback
phenomenology, but did not address the effect of stratification.
This is one of the important additions of the more recent work of
\cite{BKKR10}, of which we report here the main highlights.

\section{DNS model and analysis}

DNS of forced turbulence
were performed in a cubic computational domain of size $L^3$.
For an isothermal equation of state and a constant vertical gravitational
acceleration $g$,
one finds an exponentially stratified density:
\begin{displaymath}
\rho=\rho_0\exp\left(-z/H_\rho \right),
\end{displaymath}
where $H_\rho=c_s^2 / g$ is the constant density scale height, $c_s$
is the isothermal sound speed and $\rho_0$ a normalisation factor.
We choose $k_1 H_\rho=1$, where $k_1 =2\pi / L$, the smallest wavenumber.
The density contrast then corresponds to $\exp 2\pi\approx 535$.

We solve the equations of compressible magneto-hydrodynamics in the form
\begin{displaymath}
\rho{D\bm{U}\over D t}=\bm{J}\times\bm{B}-c_s^2\nabla\ln\rho+\nabla\cdot(2\nu\rho\SSSS)
+\rho(\bm{f}+\bm{g}),
\end{displaymath}
\begin{displaymath}
{\partial\bm{A}\over\partial t}=\bm{U}\times\bm{B}+\eta\nabla^2\bm{A},
\end{displaymath}
\begin{displaymath}
{\partial\rho\over\partial t}=-\nabla\cdot\rho\bm{U},
\end{displaymath}
where $\nu$ and $\eta$ are kinematic viscosity and magnetic diffusivity,
respectively. Furthermore, $\bm{B}=\bm{B}_0+\nabla\times\bm{A}$ is the
magnetic field consisting
of a uniform mean field, $\bm{B}_0=(0,B_0,0)$, and a nonuniform part
that is represented in terms of the magnetic vector potential $\bm{A}$,
$\bm{J}=\nabla\times\bm{B}/\mu_0$ is the current density, and
${\sf S}_{ij}=\frac{1}{2}(U_{i,j}+U_{j,i})-\frac{1}{3}\delta_{ij}\nabla\cdot\bm{U}$
is the traceless rate of strain tensor, where commas denote
partial differentiation.
The turbulence is driven with a forcing function $\bm{f}$ that consists of
random plane non-polarized waves with an average wavenumber $k_{\rm f}=5\,k_1$.
The forcing strength is arranged such that the turbulent rms velocity,
$u_{\rm rms}=\left\langle\bm{u}^2\right\rangle^{1/2}$, is around $0.1\,c_s$.
This value is small enough so that compressibility effects are weak.

In order to characterize our simulations, a set of dimensionless parameters is defined.
The Reynolds number is given by ${\rm Re}=u_{\rm rms}/\nu k_{\rm f}$
and is of the order 120 in the simulations.
The magnetic Prandtl number is $\Pm=\nu/\eta$, although in reality this
parameter is much smaller than unity;
we choose here $\Pm=0.5-8$ in order to achieve higher values for the magnetic Reynolds number.
The equipartition field strength $B_{\rm eq}$ is defined as a function
of $z$ and the imposed fields
are normalised against $B_{\rm eq0}$, the equipartition strength in the middle of the domain:
\begin{displaymath}
B_{\rm eq}\left(z\right)=\left(\mu_0\overline{\rho\bm{u}^2}\right)^{1/2},\quad
B_{\rm eq0} =\left(\mu_0\rho_0\right)^{1/2}u_{\rm rms}.
\end{displaymath}

The boundary conditions are stress-free perfect conductors at the top and bottom of the domain,
periodicity in the horizontal direction. The simulations are performed with the {\sc Pencil Code}%
\footnote{{\tt http://pencil-code.googlecode.com}},
which uses sixth-order explicit finite differences in space and a
third-order accurate time stepping method \citep{BD02}.

The contribution to the mean momentum density by the fluctuations is
\begin{displaymath}
\overline{\Pi}_{ij}^f=\bar{\rho}\overline{u_iu_j}+\frac{1}{2}\delta_{ij}\overline{b^2}-\overline{b_ib_j},
\end{displaymath}
where the overbars now indicate horizontal averages.
The influence of the mean magnetic field can be found by subtracting the
contributions that are present in the absence of a uniform background
magnetic field and can be modelled by the following ansatz (RK07)
\begin{displaymath}
\overline{\Pi}_{ij}^{f,\overline{B}}-\overline{\Pi}_{ij}^{f,\overline{0}}=-\left(\frac{1}{2}\delta_{ij}q_p+e_ie_jq_e\right)\overline{\bm{B}}^2+q_s\overline{B}_i\overline{B}_j,
\end{displaymath}
which will then appear in the effective mean Lorentz force
\begin{displaymath}
\rho \bm{F}_i^M=-\nabla_j\left(\delta_{ij}\overline{\bm{B}}^2
+\overline{B}_i\overline{B}_j+
\overline{\Pi}_{ij}^{f,\overline{B}}
-\overline{\Pi}_{ij}^{f,\overline{0}}\right).
\end{displaymath}
Thus the total effective magnetic pressure of the mean field is given by $1-q_p\left(\overline{B}\right)\overline{\bm{B}}^2$
As $\overline{\bm{B}}\approx\left(0,\overline{B},0\right)$, and assuming no small-scale dynamo, we can determine $q_p$ from

\begin{displaymath}
\rho \left(\overline{u_x^2}-\overline{u_{0x}^2}\right)+\frac{1}{2}\overline{\bm{b}^2}-\overline{b_x^2}=-\frac{1}{2}q_p\overline{\bm{B}}^2.
\end{displaymath}

\section{Results}

\begin{figure}\begin{center}
\includegraphics[width=0.49\textwidth]{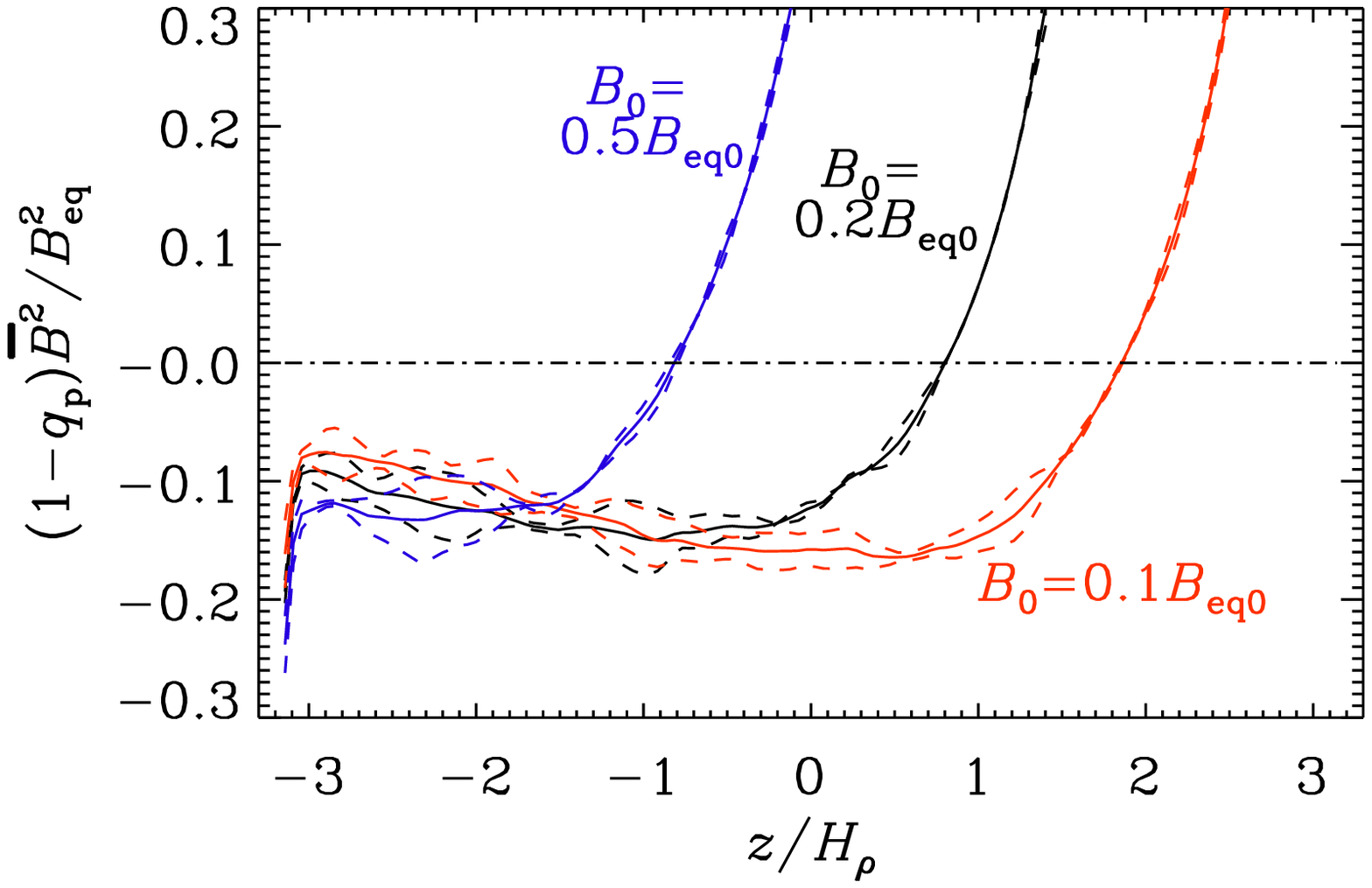}
\includegraphics[width=0.49\textwidth]{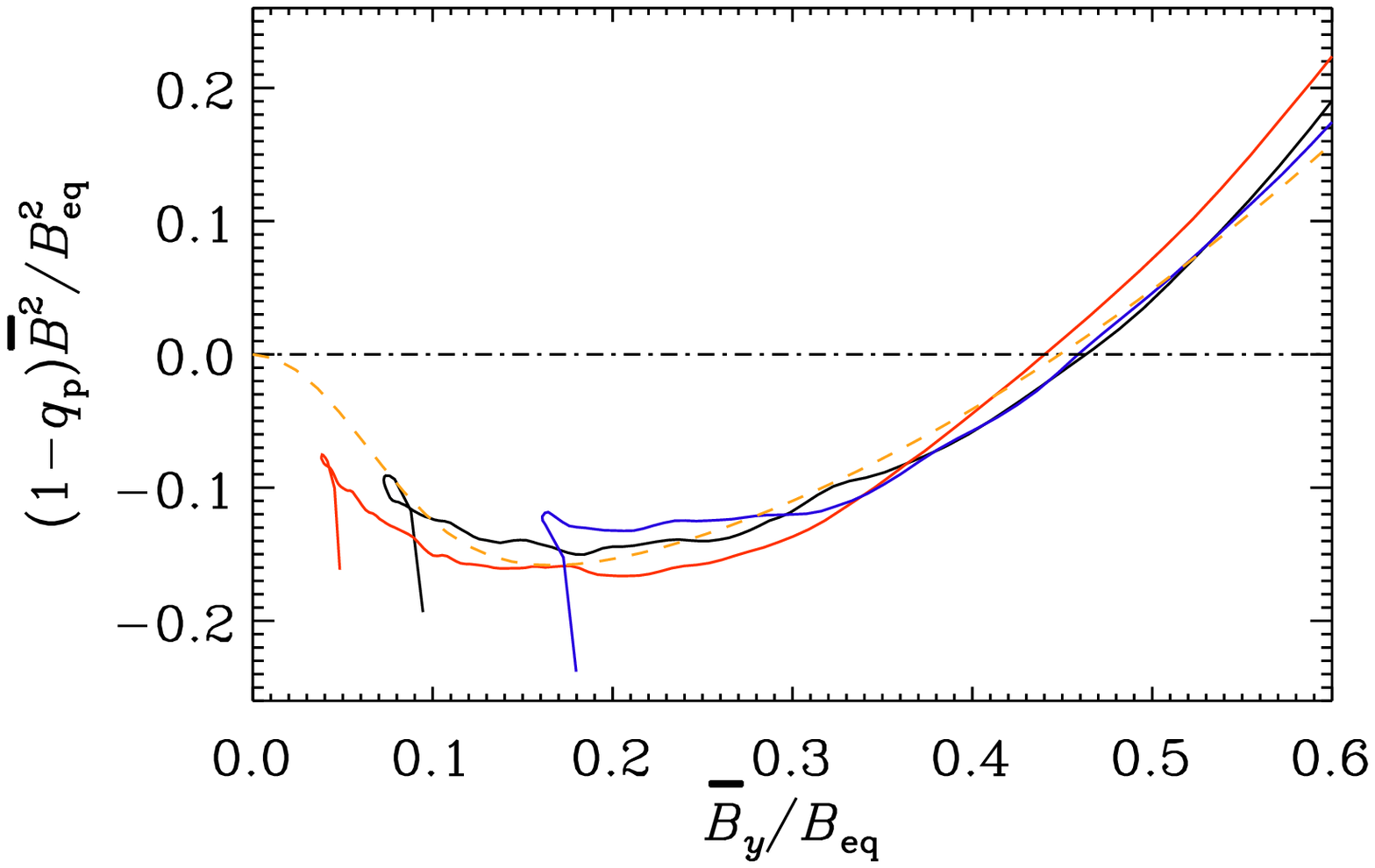}
\end{center}\caption[]{
Normalized effective mean magnetic pressure for $B_0=0.1 B_{\rm eq0}$, $B_0=0.2 B_{\rm eq0}$, and $B_0=0.5 B_{\rm eq0}$ using $\Rey=120$.
{\it Left}: as a function of depth
{\it Right}: as function of the local value of the ratio of $B_0/B_{\rm eq}(z)$.
Note that the curves for the different imposed field strengths
collapse onto a single dependence and agree very well with the fit by BKR (dashed line).
Adapted from \cite{BKKR10}.
}\label{pxyaver_compp}
\end{figure}

\begin{figure}\begin{center}
\includegraphics[width=0.52\textwidth]{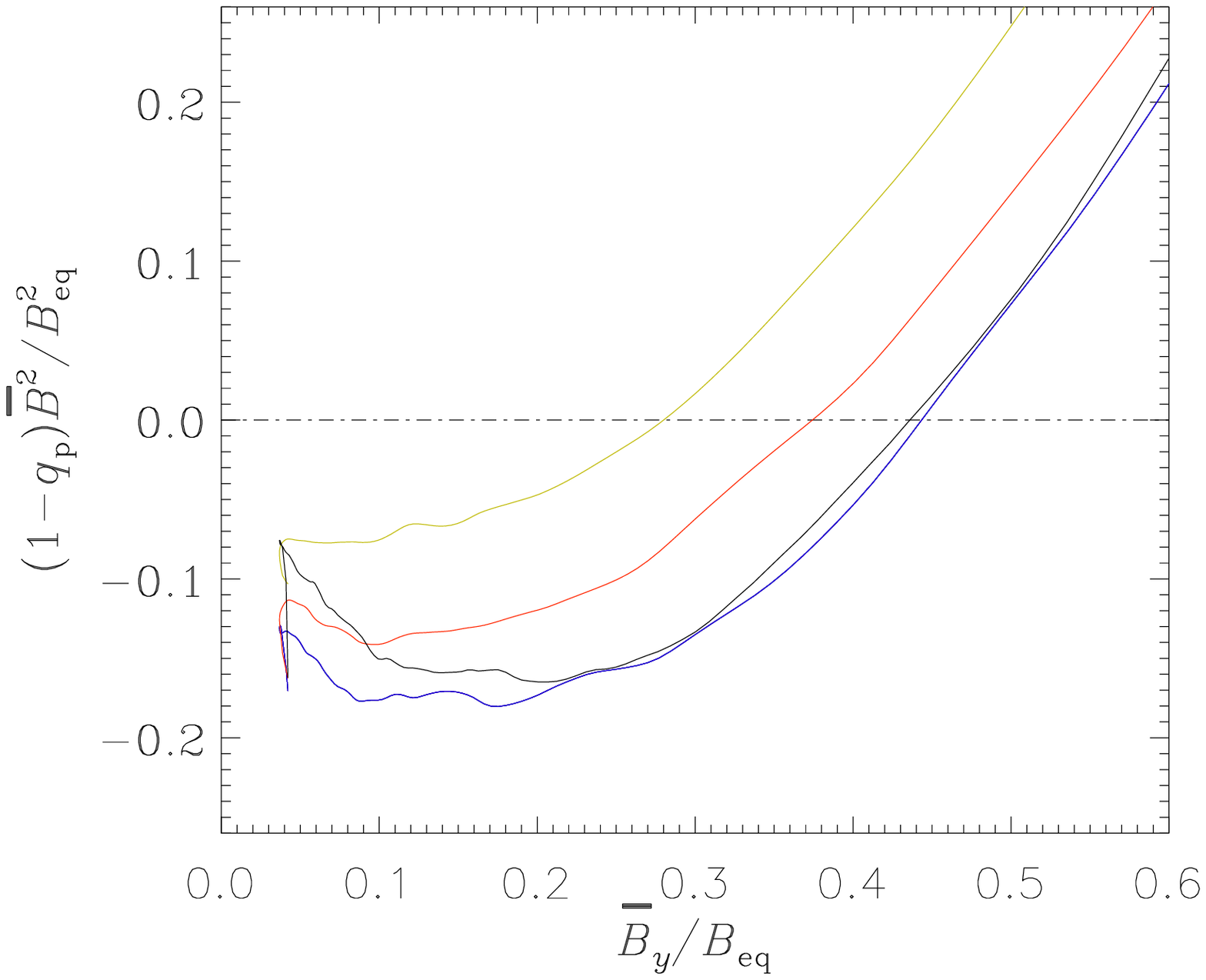}
\includegraphics[width=0.46\textwidth]{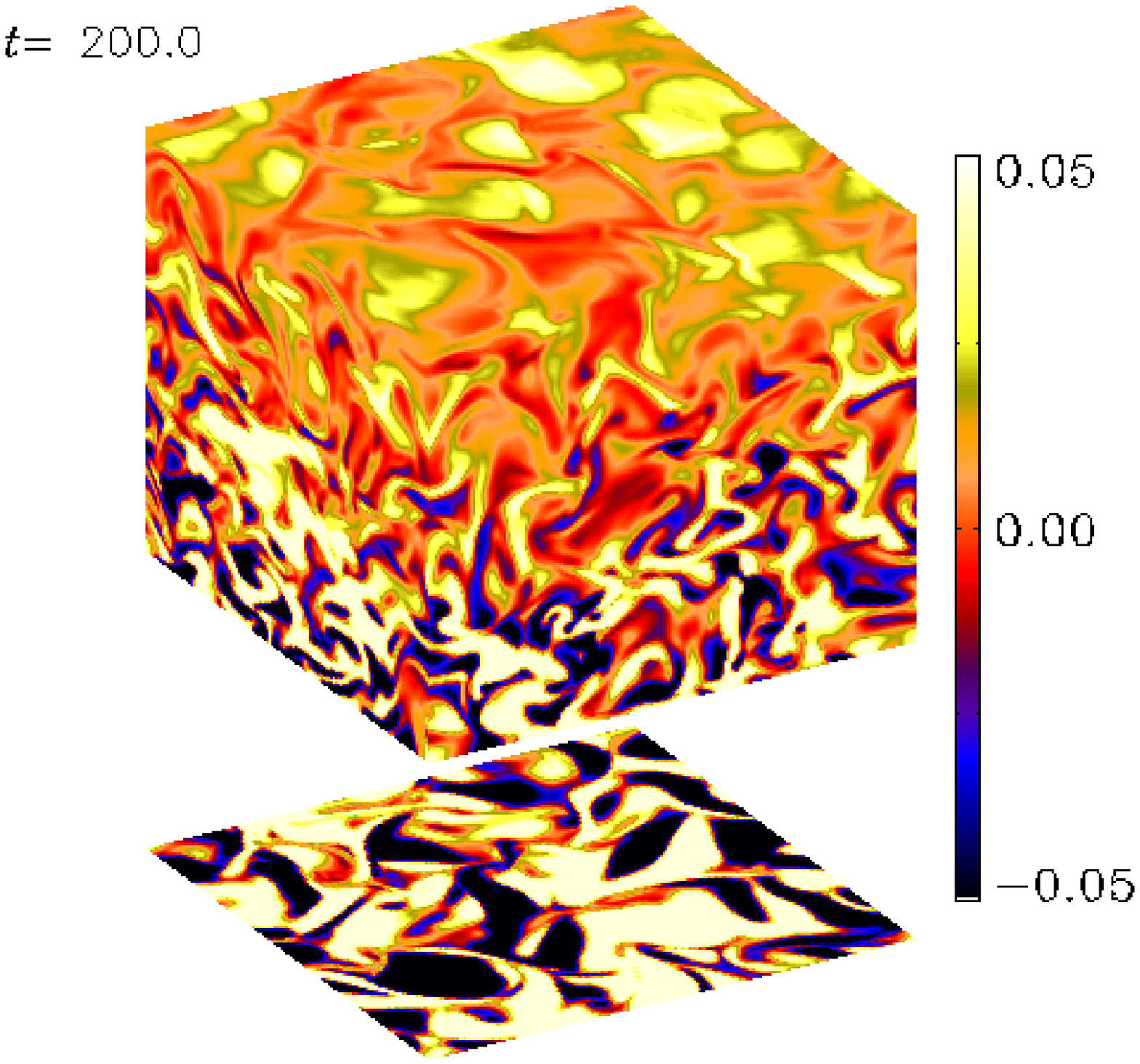}
\end{center}\caption[]{
{\it Left}:
Normalized effective mean magnetic pressure as a function of $B_0/B_{\rm eq}(z)$
for $B_0=0.1 B_{\rm eq0}$, and varying magnetic Prandtl number,$\Pm=0.05$ (black), $2$ (blue), $4$ (red) and $8$ (yellow).
{\it Right}:
Visualisation of $B_y-B_0$ on the periphery of the computational domain $B_0=0.1 B_{\rm eq0}$, and $\Pm=2$.
}\label{me_ef_ma_pr_prandtlsweep}\end{figure}

We compare the effective magnetic pressure of the mean field with the
turbulent kinetic energy density (Fig.~\ref{pxyaver_compp}a) and see
that the resulting contribution of stratified turbulence to the effective
magnetic pressure is negative in a large part of the domain.
Plotting the ratio of effective magnetic pressure to kinetic energy
density as a function of the imposed uniform horizontal magnetic field
divided by the equipartition field strength (Fig.~\ref{pxyaver_compp}b),
collapses the observations from simulations with different imposed field
strengths into a single dependence of $q_p$ on $\overline{B_y}/B_{\rm eq}$.
This result agrees with analytical calculations by RK07 and a fit
based on simulations by BKR.
The effect is fairly robust under an increase of the magnetic Prandtl
number (Fig.~\ref{me_ef_ma_pr_prandtlsweep}a).
While the reduced effective magnetic pressure is observed, the
formation of local magnetic field concentrations as observed
in mean-field simulations, has not yet been found in DNS
(Fig.~\ref{me_ef_ma_pr_prandtlsweep}b).

\section{Discussion}

The DNS have shown that for an isothermal atmosphere with strong density stratification, the turbulent pressure is decreased
due to a negative feedback from magnetic fluctuation generation, resulting in a negative effective mean magnetic pressure.
The dependence on the ratio of imposed field to local equipartition field
agrees with results obtained from analytic theory (RK07) and direct
numerical simulations (BKR). The results are robust when changing the
strength of the imposed field and the magnetic Prandtl number.

However, the simulations do not show any obvious signs  of a large-scale
instability that should result in magnetic flux concentrations,
as was expected from mean-field calculations. A possible explanation for this discrepancy
could be the simplicity of the ansatz for the effective Lorentz force.
Indeed, early simulations of \cite{Tao98} produced clear signs of
flux separation into magnetised and unmagnetised regions in convection
simulations at large aspect ratio.
An increase of scale separation might alleviate the effect of higher
order terms.
Future work should incorporate this increase as well as a wider scan of
the parameter regime and ultimately the inclusion of radiative transfer,
as was done in simulations of \cite{Kiti10}, which showed flux
concentrations in the presence of a vertical field.

\acknowledgments

We acknowledge the allocation of computing resources provided by the
Swedish National Allocations Committee at the Center for
Parallel Computers at the Royal Institute of Technology in
Stockholm and the National Supercomputer Centers in Link\"oping
as well as the Norwegian National Allocations Committee at the
Bergen Center for Computational Science.
This work was supported in part by
the European Research Council under the AstroDyn Research Project No.\ 227952
and the Swedish Research Council Grant No.\ 621-2007-4064.
NK and IR thank NORDITA for hospitality and support during their visits.

\newcommand{\yjour}[4]{ #1, {#2}, {#3}, #4}
\newcommand{\yana}[3]{ #1, {A\&A,} {#2}, #3}
\newcommand{\ysph}[3]{ #1, {Solar Phys.,} {#2}, #3}
\newcommand{\yan}[3]{ #1, {Astron.\ Nachr.,} {#2}, #3}
\newcommand{\yapj}[3]{ #1, {ApJ,} {#2}, #3}
\newcommand{\sapj}[1]{ #1, {ApJ,} submitted}
\newcommand{\ypfb}[3]{ #1, {Phys.\ Fluids B,} {#2}, #3}
\newcommand{\ypre}[3]{ #1, {Phys.\ Rev.\ E,} {#2}, #3}
\newcommand{\yjetp}[3]{ #1, {Sov.\ Phys.\ JETP,} {#2}, #3}
\newcommand{\ymn}[3]{ #1, {MNRAS,} {#2}, #3}
\newcommand{\yssr}[3]{ #1, {Spa. Sci. Rev.,} {#2}, #3}


\begin{thebibliography}{}

\bibitem[Brandenburg(2005)]{B05}
Brandenburg, A.\yapj{2005}{625}{539}

\bibitem[Brandenburg \& Dobler(2002)]{BD02}
Brandenburg, A., \& Dobler, W.\yjour{2002}{Comp. Phys. Comm.}{147}{471}

\bibitem[Brandenburg \& Subramanian(2005)]{BS05}
Brandenburg, A., \& Subramanian, K.\yjour{2005}{Phys.\ Rep.}{417}{1}

\bibitem[Brandenburg et al.(2010a)]{BKR10}
Brandenburg, A., Kleeorin, N., \& Rogachevskii, I.\yan{2010}{331}{5} (BKR)

\bibitem[Brandenburg et al.(2010b)]{BKKR10}
Brandenburg, A., Kemel, K., Kleeorin, N., \& Rogachevskii, I. 2010, arXiv:1005.5700

\bibitem[Cattaneo et al.(2006)]{Cat06}
Cattaneo, F., Brummell, N. H., Cline, K. S.\ymn{2006}{365}{727}

\bibitem[Fan(2001)]{Fan01}
Fan, Y.\yapj{2001}{546}{509}

\bibitem[Kitchatinov \& Mazur(2000)]{KM00}
Kitchatinov, L.L., \& Mazur, M.V.\ysph{2000}{191}{325}

\bibitem[Kitiashvili et al.(2010)]{Kiti10}
Kitiashvili, I. N., Kosovichev, A. G., Wray, A. A., \& Mansour, N. N.\yapj{2010}{719}{307}

\bibitem[Kleeorin et al.(1990)]{KRR90}
Kleeorin, N.I., Rogachevskii, I.V., \& Ruzmaikin, A.A.\yjetp{1990}{70}{878}

\bibitem[Nordlund et al.(1992)]{Nor92}
Nordlund, \AA., Brandenburg, A., Jennings, R. L., Rieutord, M.,
Ruokolainen, J., Stein, R. F., \& Tuominen, I.\yapj{1992}{392}{647}

\bibitem[Parker(2009)]{Par09}
Parker, E. N.\yssr{2009}{144}{15}

\bibitem[Rempel et al.(2009)]{Remp09}
Rempel, M., Sch\"{u}ssler, M., \& Kn\"{o}lker, M.\yapj{2009}{691}{640}

\bibitem[Rogachevskii \& Kleeorin(2007)]{RK07}
Rogachevskii, I., \& Kleeorin, N.\ypre{2007}{76}{056307}(RK07)

\bibitem[Tao et al.(1998)]{Tao98}
Tao, L., Weiss, N. O., Brownjohn, D. P., \& Proctor, M. R. E.\yapj{1998}{496}{L39}

\bibitem[Tobias et al.(1998)]{Tob98}
Tobias, S. M., Brummell, N. H., Clune, T. L., \& Toomre, J.\yapj{1998}{502}{L177}

\end{thebibliography}
\end{document}